%Paper: hep-th/9406035
%From: sonoda@physics.ucla.edu (Sonoda)
%Date: Tue, 7 Jun 94 14:21:43 -0700

% contribution to International Colloquim on
% Modern Quantum Field Theory, 5--11 January 1994
% June 1994
\headline={\ifnum\pageno=1\firstheadline\else
\ifodd\pageno\rightheadline \else\leftheadline\fi\fi}
\def\firstheadline{\hfil}
\def\rightheadline{\hfil}
\def\leftheadline{\hfil}
	\footline={\ifnum\pageno=1\firstfootline\else\otherfootline\fi}
\def\firstfootline{\rm\hss\folio\hss}
\def\otherfootline{\hfil}

\font\twelvebf=cmbx10 scaled\magstep 1
\font\twelverm=cmr10 scaled\magstep 1
\font\twelveit=cmti10 scaled\magstep 1

\font\elevenrm=cmr10 scaled\magstephalf

\font\tenbf=cmbx10
\font\tenrm=cmr10
\font\tenit=cmti10

\font\ninerm=cmr9

\parindent=1.5pc
\hsize=6.0truein
\vsize=8.5truein
\nopagenumbers

\overfullrule=0pt
\hfill
{\elevenrm UCLA/94/TEP/20}
\vglue 0.6cm
\centerline{\tenbf GEOMETRICAL EXPRESSION FOR}
\baselineskip=16pt
\centerline{\tenbf SHORT-DISTANCE
SINGULARITIES IN FIELD THEORY$^\star$}
\footnote{}{\ninerm $^\star$
Based upon a talk given at the International
Colloquim on Modern Quantum Field Theory,
5--11 January 1994, Bombay, India.
This work was supported in part
by the U.S. Department
of Energy, under Contract DE-AT03-88ER 40384 Mod A006 Task C.}

\vglue 0.8cm
\centerline{\tenrm HIDENORI SONODA}
\baselineskip=13pt
\centerline{\tenit Department of Physics and Astronomy,
UCLA}
\baselineskip=12pt
\centerline{\tenit Los Angeles, CA 90024-1547, USA}
\vglue 0.8cm
\centerline{\tenrm ABSTRACT}
\vglue 0.3cm
{\rightskip=3pc
\leftskip=3pc
\tenrm\baselineskip=12pt\noindent
We consider the linear space of composite fields as an infinite
dimensional vector bundle over the theory space
whose coordinates are simply
the parameters of a renormalized field theory.
We discuss a geometrical
expression for the short distance singularities
of the composite fields in terms of beta functions,
anomalous dimensions, and a connection.
\vglue 0.6cm}
% begin definitions
\def\dt{{d\over dt}~}
\def\O{{\cal O}}
\def\vev#1{\left\langle #1 \right\rangle_g}
\def\L{{\cal L}}
% end definitions
\twelverm
\baselineskip=14pt
\leftline{\twelvebf 1. Introduction}
\vglue 0.4cm
The idea of theory space was introduced
to understand continuum limits in field theory.$^1$ Though we have
gained much qualitative understanding of field theory
(e.g., renormalization group (RG) flows, fixed points, universality),
very few quantitative results have been obtained so far.
A notable exception is the c-theorem on the monotonic
decrease of the c-function along the RG flows
on the space of two-dimensional field theories.$^2$

It is important to understand the theory space quantitatively,
since the theory space is the natural framework
in which we discuss field theory.  For example,
the continuum limits of field theories can be defined
non-perturbatively as
the image of an infinite number of RG transformations
on the theory space.

Another reason we should study the theory space
is related to a yet unknown non-perturbative
formulation of string theory.  It has been advocated
for long that the string action should be a function
on the space of all possible two dimensional
field theories.  We hardly know anything about
the space.

The theory space we will discuss below
is more restricted than the full theory space that
includes all possible field theories with a UV cutoff.
We will only consider the space of renormalizable field
theories.  Then, given a renormalizable theory with
$N$ parameters, we obtain an $N$-dimensional theory
space.  We will first clarify the geometrical objects
on the theory space and next show that certain short
distance singularities adopt a geometrical interpretation.

\vfil\eject
%\vglue 0.6cm
\twelverm
\leftline{\twelvebf 2. Theory Space}
\vglue 0.4cm
We consider a renormalized field theory in $D$-dimensional
euclidean space with parameters $g^i (i=1,...,N)$.  Let
the origin $g^i = 0$ be an UV fixed point.  We regard $g$'s
as local coordinates of a finite dimensional theory space.
The parameters satisfy the renormalization group equations
$$
\dt g^i = \beta^i (g) .\eqno(1)
$$
The beta functions $\beta^i$ form a vector field on the
theory space.

Let $\O_i$ be a scalar composite field conjugate to $g^i$:
the field $\O_i$ generates a deformation of the theory
in the $g^i$ direction.  The conjugate fields make a basis
of the tangent vector bundle over the theory space, since
the fields transform as
$$
\O_i \to \O'_i = {\partial g^j \over \partial {g'}^i} \O_j \eqno(2)
$$
under an arbitrary coordinate change $g \to g'$.

We introduce a basis of composite fields $\{ \Phi_a \}_g$ at
each point $g$ on the theory space.  Since the composite
fields make a linear space at every $g$, they form
an infinite dimensional vector bundle over the theory space.
We let $\Gamma_a^{~b} (g)$ be the full scale dimension of
the composite fields:
$$
\dt \Phi_a = \Gamma_a^{~b} (g) \Phi_b .\eqno(3)
$$
The choice of a basis is not unique, and
we can transform the basis as
$$
\Phi_a \to \Phi'_a = N_a^{~b} (g) \Phi_b ,\eqno(4)
$$
where $N(g)$ is an invertible matrix.
For an arbitrary choice of $n$ {\twelvebf different} points
$r_k (k=1,...,n)$ in space, the correlation function
$\vev{\Phi_{a_1} (r_1) ... \Phi_{a_n} (r_n)}$ make
a basis of rank-$n$ tensor fields over the theory space.

We now consider the operator product expansion (OPE)
$$
\O_i (r) \Phi_a (0) = {1 \over {\rm vol} (S^{D-1})}~(C_i)_a^{~b} (r;g) \Phi_b
(0)
+ {\rm o} \left( {1 \over r^D} \right) ,\eqno(5)
$$
where $C_i (r;g)$ denotes the part of the OPE coefficient which
cannot be integrated over volume.
We define a matrix by
$$
H_i (g) \equiv C_i (r=1;g) .\eqno(6)
$$
This is a tensor over the theory space, since
it transforms as
$$
H_i (g) \to N(g) H_i (g) N(g)^{-1} \eqno(7)
$$
under Eq.~(4).

\vfil\eject
%\vglue 0.6cm
\leftline{\twelvebf 3. Geometrical Expression for Short Distance Singularities}
\vglue 0.4cm
The following geometrical expression for the
tensor $H_i$, defined by Eq.~(6), has been derived$^{3,4}$:
$$
H_i (g) = {\partial \Psi (g) \over \partial g^i} + [c_i (g), \Psi (g)] +
\beta^j (g) \Omega_{ji} (g) .
\eqno(8)
$$
where
$$
\Psi \equiv \Gamma + \beta^i c_i ,\quad \Omega_{ji} \equiv \partial_j c_i -
\partial_i c_j
+ [c_j, c_i] .\eqno(9)
$$
It is easy to check that $\Psi (g)$ has the same transformation property
as $H_i$ under Eq.~(4).
The tensor $\Omega_{ji}$ is the curvature of the connection $c_i$, and it has
a field theoretic expression:
$$
\eqalign{
&\quad \Omega_{ji} \vev{\Phi} \cr
&= \int_{r \le 1} d^D r ~{\rm F.P.}_r \int_{r' \le 1} d^D r'~
\vev{ \left( \O_i (r) \left( \O_j (r') - {1 \over {\rm vol} (S^{D-1})} C_j (r')
\right)
- ( r \leftrightarrow r' ) \right) \Phi (0)}^c ,\cr}\eqno(10)
$$
where ${\rm F.P.}_r$ stands for taking the integrable part in $r$.
We will not repeat the derivation of Eqs.~(8) and (10) here.  They have been
derived on the basis
of an assumption that $\O_i$ generates deformation of the theory
under a change of $g^i$.  Then the consistency with the RG gives
the above expressions.

If we restrict our attention to the conjugate fields alone, then
the matrix $\Gamma$ is given by
$$
\Gamma_i^{~j} = D \delta_i^j - {\partial \beta^j \over \partial g^i} .\eqno(11)
$$
Hence,
$$
\Psi_i^j = D \delta_i^j - {\partial \beta^j \over \partial g^i} + \beta^k
c_{k,i}^{~~j}
= D \delta_i^j - \nabla_i \beta^j ,\eqno(12)
$$
and
$$
(H_i)_j^{~k} = - \nabla_i \nabla_j \beta^k + \beta^l (\Omega_{li})_j^{~k}
,\eqno(13)
$$
where $\nabla_i$ is the covariant derivative, and
we have used the symmetry of the connection
$$
(c_i)_j^{~k} = (c_j)_i^{~k} .\eqno(14)
$$
Eq.~(13) is a recent result of Dolan.$^5$
Note the symmetry
$$
(H_i)_j^{~k} = (H_j)_i^{~k} \eqno(15)
$$
due to the cyclic symmetry of the curvature:
$$
(\Omega_{li})_j^{~k} + (\Omega_{jl})_i^{~k} + (\Omega_{ij})_l^{~k} = 0
.\eqno(16)
$$
It is natural to ask if the connection $(c_i)_j^{~k}$ is the unique
Riemann connection of some metric $G_{ij}$ on the theory space.
We do not know the answer in general, though on the space
of two-dimensional conformal field theories the metric is given
by the two-point function of the conjugate fields.$^{2,6}$

The connection arises naturally in field theory by considering
infinitesimal changes of the correlation function
$\vev{\Phi_{a_1} (r_1) ... \Phi_{a_n} (r_n)}$.  The naked derivative
$\partial/\partial g^i$ simply spoils the covariance of the correlation
function under Eq.~(4).  To restore covariance we must
introduce a connection.  Then the covariant derivative
$$
\partial_i \vev{\Phi_{a_1} (r_1) ... \Phi_{a_n} (r_n)}
+ \sum_{k=1}^n (c_i)_{a_k}^{~b} \vev{\Phi_{a_1} (r_1) ... \Phi_b (r_k) ...
\Phi_{a_n} (r_n)}
$$
is given by an integral of the conjugate field $\O_i$ with the UV singularities
properly subtracted.  Thus, the connection arises as finite counterterms.$^3$

\vglue 0.6cm
\leftline{\twelvebf 4. Examples}
\vglue 0.4cm
We consider two examples here using dimensional regularization.
\vglue 0.3cm
\leftline{\twelveit 4.1. $(\phi^4)_4$ Theory}
\vglue 0.15cm
The lagrangian is given by
$$
\L = {1 \over 2} \partial_\mu \phi_0 \partial_\mu \phi_0 + {Z_m m^2 \over 2}
\phi_0^2
+ {Z_\lambda \lambda \over 4!} \phi_0^4 ,\eqno(17)
$$
where $\phi_0$ is a bare field.
Hence, the conjugate fields are
$$
\eqalign{
\O_\lambda &= {\partial \L\over \partial \lambda}\Big|_{m^2,\phi_0} =
{\partial (Z_\lambda \lambda) \over \partial \lambda} {\phi_0^4 \over 4!}
+ {\partial Z_m \over \partial \lambda} m^2 {\phi_0^2 \over 2} ,\cr
\O_m &= {\partial \L\over \partial m^2}\Big|_{\lambda,\phi_0} =
Z_m {\phi_0^2 \over 2} .\cr}\eqno(18)
$$
These fields are renormalized.\footnote{*}{\ninerm
In fact $\O_\lambda$ may still need counterterms proportional to
$\partial^2 {\phi_0^2 \over 2}$ and $m^4 {\bf 1}$.
Likewise $\O_m$ may need a term proportional to
$m^2 {\bf 1}$.}$^7$

We can choose the parameters $\lambda, m^2$ such
that the following gauge conditions are satisfied:
$$
(c_\lambda)_\lambda^{~\lambda} (\lambda) = (c_\lambda)_m^{~m} (\lambda) = 0
.\eqno(19)
$$
This may amount to choosing a non-minimal subtraction scheme
for $Z_\lambda, Z_m$.  In the above gauge, we find
$$
(c_\lambda)_\lambda^{~m} = m^2 (\Omega_{m\lambda})_{\lambda}^{~m} ,\eqno(20)
$$
which turns out to be nonvanishing.$^3$  Eq.~(8) (or (13)) implies
$$
\eqalign{(H_m)_\lambda^{~m} &= (H_\lambda)_m^{~m} = - \beta_m' \cr
(H_\lambda)_\lambda^{~\lambda} &= - \beta_\lambda'' (\lambda) , \cr}\eqno(21)
$$
where $\beta_\lambda$ is the beta function, and $\beta_m$ is the
anomalous dimension of the mass parameter.
More details can be found elsewhere.$^3$

\vglue 0.3cm
\leftline{\twelveit 4.2. $O(N)$ Nonlinear Sigma Model in $D=2$}
\vglue 0.15cm
The lagrangian is given by
$$
\L = {\partial_\mu \Phi_0^I \partial_\mu \Phi_0^I \over 2 Z_g g} ,\eqno(22)
$$
where $\Phi_0^I (I=1,...,N)$ is a bare field normalized by $\Phi_0^I \Phi_0^I =
1$,
and $g$ is a renormalized temperature.
The field conjugate to $g$ is
$$
\O_E = {\partial \L \over \partial g}\Big|_{\Phi_0} = {\partial \over \partial
g}
\left({1 \over Z_g g}\right)
{1 \over 2} \partial_\mu \Phi_0^I \partial_\mu \Phi_0^I .\eqno(23)
$$
This is a renormalized field with anomalous dimension
$- \beta_E'(g)$, where $\beta_E(g)$ is the beta function.

Let us consider a dimensionless scalar field $\Phi^{I_1 ... I_n}$ which is a
traceless
rank-n symmetric tensor under $O(N)$.  Let $\gamma_n (g)$ be its anomalous
dimension.  Then,
$$
\O_E (r) \Phi^{I_1 ... I_n} (0) = {1 \over 2\pi} (C_E)_n^{~n} (r;g) \Phi^{I_1
... I_n} (0)
+ {\rm o} \left( {1 \over r^2} \right) ,\eqno(24)
$$
where the OPE coefficient is given by
$$
(C_E)_n^{~n} (r;g) = {\beta_E (g(\ln r)) \over \beta_E (g)} ~(H_E)_n^{~n}
(g(\ln r)) ,\eqno(25)
$$
and $g(t)$ is the running parameter satisfying
$$
\dt g(t) = \beta_E (g(t)) \quad (g(0) = g) .\eqno(26)
$$
Eq. (8) implies that
$$
(H_E)_n^{~n} (g) = {d \over dg} \left( \gamma_n + \beta_E (c_E)_n^{~n} \right)
.\eqno(27)
$$
We can change the normalization of the field $\Phi^{I_1 ... I_n}$ so that
$$
(c_E)_n^{~n} (g) = 0 .\eqno(28)
$$
Then, we simply get
$$
(H_E)_n^{~n} (g) = \gamma_n'(g)  .\eqno(29)
$$
More details will be provided elsewhere.$^8$

\vglue 0.6cm
\leftline{\twelvebf 5. Conclusion}
\vglue 0.4cm
The results we have presented in the above are modest:
the short distance singularities in the products of the conjugate
fields and composite fields can be expressed geometrically
in terms of beta functions, anomalous dimensions, and a connection.
We would like to think that
this is an indication that field theory awaits geometrical interpretation.

A more ambitious program is to consider the full theory space, i.e.,
the space of all possible field theories with a UV cutoff.
In this case, however, we may not obtain any interesting
{\twelvebf local} geometry.  As we have remarked at the end
of sect.~2, the connection arises as finite counterterms
necessary for UV subtracted integrals.
The cutoff theories do not need any counterterm,
so we expect that a scheme exists in which the connection
vanishes locally on the theory space.  But it is possible to
find interesting {\twelvebf global}
geometry with the flat connection.

We do not have any a priori reason that field theory
should be interpreted geometrically, but we think
that any geometrical structure
of theory space is worth studying for its own sake.
\vfil
\vglue 0.6cm
\leftline{\twelvebf References}
\medskip
\itemitem{1.} K.~G.~Wilson and J.~Kogut, {\twelveit Phys. Rept.}~{\twelvebf
C12}(1974)75.
\itemitem{2.} A.~B.~Zamolodchikov, {\twelveit JETP Lett.}~{\twelvebf
43}(1986)730.
\itemitem{3.} H.~Sonoda, {\twelveit Nucl. Phys.}~{\twelvebf B383}(1992)172;
{\twelveit Nucl. Phys.}~{\twelvebf B394}(1993)302.
\itemitem{4.} H.~Sonoda, ``Connection on the theory space,'' hep-th/9306119.
\itemitem{5.} B.~Dolan, ``Co-variant derivatives and the renormalization
group,''
hep-th/9403070.
\itemitem{6.} K.~Ranganathan, H.~Sonoda, and B.~Zwiebach,
{\twelveit Nucl. Phys.}~{\twelvebf B414}(1994)405.
\itemitem{7.} L.~Brown, {\twelveit Ann. Phys.}~{\twelvebf 126}(1980)135.
\itemitem{8.} H.~Sonoda and W.-C.~Su, ``Operator Product Expansions
in the Two-Dimensional O(N) Non-Linear Sigma Model,'' hep-th/9406007.

\bye